\RequirePackage{fix-cm}
\documentclass[twocolumn]{svjour3}          
\smartqed  
\usepackage{graphicx}
\begin{document}

\title{The Chaplygin sleigh with friction moving due to periodic oscillations of an
internal mass
\thanks{Russian Science Foundation, Grant no. 15-12-20035}
}


\author{Ivan A. Bizyaev         \and
       Alexey V. Borisov         \and
        Sergey P.~Kuznetsov 
}

\institute{Ivan A. Bizyaev \at
              Udmurt State University, Universitetskay 1, Izhevsk, 426034, Russian Federation \\
              \email{bizaev{\_}90@mail.ru}           
           \and
           Alexey V. Borisov \at
              Udmurt State University, Universitetskay 1, Izhevsk, 426034, Russian Federation \\
              \email{borisov@rcd.ru}
              \and
           Sergey P.~Kuznetsov  (corresponding author) \at
              Udmurt State University, Universitetskay 1, Izhevsk, 426034, Russian Federation, \\
              Kotel'nikov's Institute of Radio-Engineering and Electronics of RAS,
              Saratov Branch, Zelenaya 38, Saratov, Russian Federation, \\
              \email{spkuz@yandex.ru}
}

\date{Received: date / Accepted: date}

\maketitle

\begin{abstract}
For a Chaplygin sleigh moving in the presence of weak friction, we present
and investigate two mechanisms of arising acceleration due to oscillations
of an internal mass. In certain parameter regions, the mechanism induced by
small oscillations determines acceleration which is on average
one-directional. The role of friction is that the velocity reached in the
process of the acceleration is stabilized at a certain level. The second
mechanism is due to the effect of parametric excitation of oscillations,
when the internal moving particle is comparable in mass with the main platform,
and, as occurs, a necessary condition for the acceleration is presence of friction. The
parametric instability and the resulting acceleration of the sleigh turn out
to be bounded if the line of oscillations of the moving mass is displaced
from the center of mass. The steady-state regime of motion is in many cases
associated with a chaotic attractor; accordingly, the motion of the sleigh
turns out to be similar to the process of random walk on a plane.

\keywords{Nonholonomic mechanics \and
Chaplygin sleigh \and
parametric oscillator \and
strange attractor \and
Lyapunov exponent \and
chaotic dynamics}
\end{abstract}

\section*{Introduction}
Nonholonomic mechanical systems, where algebraic relations (constraints)
imposed on dynamical variables are given by equations involving generalized
velocities, arise in the context of many problems of practical importance,
for example, in considering mobile devices, and have been a subject of many
classical studies \cite{1,2,3,4,5,6}. With basic definitions, historical aspects and the modern state of the
theory of nonholonomic dynamical systems the reader can get acquainted with
the review article \cite{6a}. Recent developments in robotics have
given a new impetus to this area of research in mechanics.
Nonholonomic
systems serve as simple models, which are attractive as compared with
alternative methods of description, which require a detailed analysis of friction
effects, and provide insight on main properties of mechanical motions
delivering efficient tools, particularly, for developing motion control
algorithms \cite{6,7,8,9,10}.

The class of nonholonomic systems is very wide and is characterized by a
hierarchy of dynamical behaviors ranging from regular motions of integrable
systems to complex dynamics of nonintegrable systems, depending on a number
of invariants and symmetries inherent in the problem \cite{11}. In particular,
systems with the above-mentioned complex dynamics include the rattleback
problem concerned with motions of a rigid body with a convex smooth surface
on a rough plane (when the principal axes determining the geometric
properties of the body and those determining the dynamical properties are
different). In such systems, mechanical energy is conserved, but there is no
preservation of phase volume, so that in the course of time evolution an
element of the volume can undergo, depending on its location, either
compression or extension. For this reason, they can exhibit phenomena which
might seem to be specific only to dissipative dynamics, like attractive
fixed points, limit cycles, and even strange attractors \cite{12,13,14}.

The Chaplygin sleigh \cite{15,16} is a well-known and well-established
paradigmatic model of nonholonomic mechanics. The Chaplygin sleigh is a
rigid body -- a platform moving on a horizontal plane under the condition
that the translational velocity at some point of the sleigh is always
oriented along a certain direction fixed relative to the platform. This
condition is a nonholonomic constraint which is imposed on the system and
which can be implemented by means of a knife edge (skate) fastened to the
sleigh or by means of a wheel pair \cite{17}. A detailed analytical study of the
problem of the sleigh motion using explicit quadratures was carried out in
the 1930s by Carath\'{e}odory \cite{16}. Depending on the position of the center
of mass relative to the knife edge, the sleigh asymptotically tends to a
rectilinear motion or to a circular motion. It is interesting that the
dynamics of the Chaplygin sleigh on an inclined plane is not integrable and
exhibits random-like asymptotic behavior depending on initial conditions
\cite{18}.

Various generalizations of the problem of the Chaplygin sleigh were
considered by many authors \cite{19,20,21,22,23,24,25,26a,26}. In particular,
attention was given to problems of the motion of the sleigh in the presence
of periodic impulse impacts \cite{19,20}, with periodic switchovers of the
nonholonomic constraint to different locations \cite{22,23}, and with periodic
transfers of the massive load placed on the sleigh \cite{24,25}. Analysis was
also made of the motion of the sleigh under the action of random forces
which model a fluctuating continuous medium \cite{26}, when, according to the
authors, the motion resembles random walks of bacterial cells with a
diffusion component.

The study of the possibility of directed motion on a plane or in a medium
induced by the presence of internal moving masses \cite{7,27,8,9,28,29,21}
is of fundamental importance from the viewpoint of robotics. In particular,
such problems are obviously interesting and insightful with respect to
objects that are described in terms of models of nonholonomic mechanics.

In \cite{24}, general equations are formulated which describe the motion of the
Chaplygin sleigh with a given periodic transfer of several internal point
masses relative to the platform. It is shown that, depending on the
parameters, the system exhibits different types of motions, including those
corresponding to strange attractors, with which chaotic (diffusion)
trajectories of the sleigh on a plane are associated. This is interpreted as
a nonholonomic analog of the well-known phenomenon of Fermi's acceleration
\cite{30,31}.

In \cite{25}, we considered a special case of the Chaplygin sleigh moving due to
a given periodic motion of a single internal point mass perpendicularly to
the axis passing through the center of mass of the platform and through the
point of application of the nonholonomic constraint. We showed a possibility
of unbounded acceleration of regular directed motion of the sleigh due to
small oscillations of the internal mass, so that the longitudinal momentum
of the sleigh increases asymptotically, in proportion to time to the $1/3$
power. Using numerical simulation, we also revealed and demonstrated
periodic, quasi-periodic and chaotic motions, which are related to
attractors corresponding to bounded variations in the velocity. Since the
equations of the system are invariant under time reversal, the presence of
attractors which correspond to the observed steady motions implies the
presence of repellers which can be found when tracking the dynamics in
backward time.

This paper, which is a further development of \cite{25}, is devoted to the study
of the motion of the Chaplygin sleigh, induced by periodic oscillations of
an internal mass, in the presence of a weak friction. In Section 1 we
formulate basic equations of a mathematical model. In Section 2 we analyze
the mechanism of acceleration that takes place under small oscillations of
the internal mass and leads to regular, on average unidirectional motion of
the sleigh; in this case the role of friction is to stabilize the velocity
of motion at a fixed level -- the smaller the friction, the higher the
velocity. Section 3 is concerned with the other mechanism of acceleration
due to the effect of parametric excitation of oscillations in a situation
where the oscillating mass constitutes a major portion of the mass of the
platform. In the case of motion of the internal mass in a straight line
passing through the center of mass, the problem reduces to a linear equation
with periodic coefficients, the increase in parametric oscillations turns
out to be unbounded and the linear dissipation (viscous friction) does not
lead to saturation. Parametric instability, and thus the degree of
acceleration, of the sleigh are bounded if the line of oscillations of the
internal mass is displaced from the center of mass. The steady-state regime
of motion corresponds to attractors of the reduced system of equations,
which in many cases are chaotic; accordingly, the motion of the sleigh
undergoing acceleration is also chaotic and similar to a diffusion process.
Section 4 discusses differences between the patterns of motions associated
with chaotic attractors in the nonholonomic model without friction and in
the model with friction. In the Conclusion (Section 5) we summarize and
discuss the results of this research.

\section{Basic equations}
We consider a mechanical system, the sleigh, the basis of which is a
platform capable of sliding without friction on a horizontal plane with a
constraint supplied at some point $R$ restricted to move exclusively in a
certain direction fixed relative to the platform that can be thought as
direction of the attached "knife-edge" (Fig.1). Additionally, a material
point of mass $m_{p}$ is placed on the sleigh that performs a predetermined
oscillatory motion on it.

Let $u_1$, $u_2 $ be the projections of the velocity of the point $R$
measured in a laboratory coordinate system \textit{OXY} onto the axes of a moving
reference frame $R\xi \eta $, and let $\omega $ be the angular velocity of
the platform. The kinetic energy of the platform, together with a moving
material point which has coordinates $\xi _p (t)$, $\eta _p (t)$ in the
reference frame attached to the platform, is
\begin{equation}
\label{eq1}
\begin{array}{l}
 T = \textstyle{1 \over 2}m_0 u_1^2 + \textstyle{1 \over 2}m_0 (u_2 +
a\omega )^2 + \textstyle{1 \over 2}J_0 \omega ^2 \\
 + \textstyle{1 \over 2}m_p (u_1 + \dot {\xi }_p - \eta _p \omega )^2 +
\textstyle{1 \over 2}m_p (u_2 + \dot {\eta }_p + \xi _p \omega )^2. \\
 \end{array}
\end{equation}
Here, $m_{0}$ is the mass of the platform, $m_{p}$ is the mass of the
particle, and $J_{0}$ is the moment of inertia of the platform relative to
its center of mass. Let
\begin{equation}
\label{eq2}
R = \textstyle{1 \over 2}c_1 u_1^2 + \textstyle{1 \over 2}c_2 u_2^2 +
\textstyle{1 \over 2}c_3 \omega ^2
\end{equation}
be the dissipative Rayleigh function , where the coefficients $c_{1,2 }$ are
responsible for friction along the axes $\xi $ and $\eta $, and the
coefficient $c_{3}$, for friction with respect to the rotational motion.

\begin{figure}
  \includegraphics[width=3.3in]{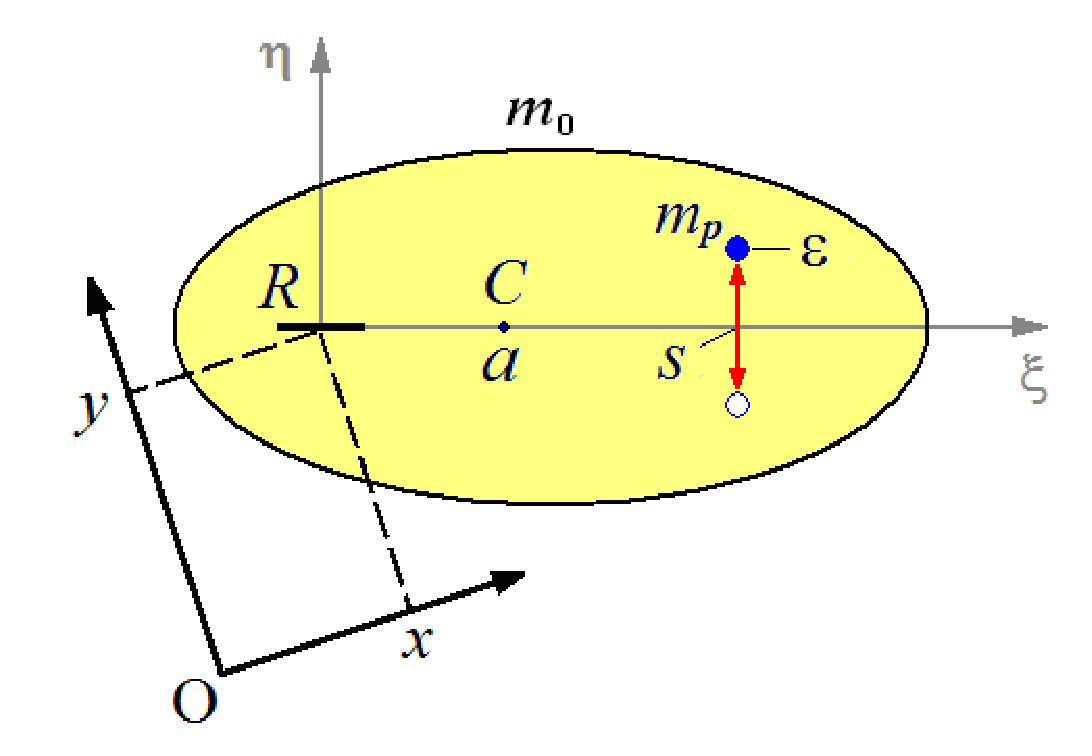}
\caption{Symbols and coordinate systems for the Chaplygin sleigh with the
oscillating internal mass $m_{p}$. The other symbols are: $m_{0}$ is the mass
of the platform, $R$ is the point of application of the nonholonomic constraint
(the position of the knife edge), and C is the center of mass of the
platform. O\textit{xy} is the laboratory reference frame, and $R\xi \eta $ is the
coordinate system attached to the platform.}
\label{fig:1}       
\end{figure}

Using these relations, we write the Lagrange equations
\begin{equation}
\label{eq3}
\begin{array}{l}
 \frac{d}{dt}\left( {\frac{\partial T}{\partial u_1 }} \right) = \omega
\frac{\partial T}{\partial u_2 } - \frac{\partial R}{\partial u_1 },\, \\
 \frac{d}{dt}\left( {\frac{\partial T}{\partial u_2 }} \right) = - \omega
\frac{\partial T}{\partial u_1 } + \lambda - \frac{\partial R}{\partial u_2
}\,,\, \\
 \frac{d}{dt}\left( {\frac{\partial T}{\partial \omega }} \right) = u_2
\frac{\partial T}{\partial u_1 } - u_1 \frac{\partial T}{\partial u_2 } -
\frac{\partial R}{\partial \omega },\,\, \\
 \end{array}
\end{equation}
where $\lambda $ is a Lagrange multiplier defined to ensure the imposed
non-holonomic constraint condition $u_{2}=0$. (In fact, it is sufficient
simply to omit the second relation since $\lambda $ does not enter the
remaining equations.)

Introducing the generalized momenta
\begin{equation}
\label{eq4}
\begin{array}{l}
 P = \left. {\frac{\partial T}{\partial u_1 }\,\,} \right|_{v_2 = 0} = m_0
u_1 + m_p (u_1 + \dot {\xi }_p - \eta _p \omega ), \\
 M = \left. {\frac{\partial T}{\partial \omega }\,\,} \right|_{v_2 = 0} =
(m_0 a^2 + J_0 + m_p (\xi _p^2 + \eta _p^2 ))\omega \\
\,\,\,\,\,\,\,\,\,\,\,\,\,\,\,\,\,\,\,\,\,\,\,\,\,\,\,\,\,\,\,\,\,\,\,\,\,\,\,\,\,\,\,\,\,\,
- m_p u_1 \eta _p + m_p(\xi _p \dot {\eta }_p - \dot {\xi }_p \eta _p ), \\
 \end{array}
\end{equation}
and assuming $m = m_0 + m_p $, $\mu = m_p / m$, $I_0 = J_0 / m$, we rewrite
the equations (\ref{eq3}) as
\begin{equation}
\label{eq5}
\begin{array}{l}
 \dot {P} = m\{a\omega + \mu [\dot {\eta }_p + (\xi _p - a)\omega ]\}\omega
- c_1 u_1 , \\
 \dot {M} = - m\{a\omega + \mu [\dot {\eta }_p + (\xi _p - a)\omega ]\}u_1 -
c_3 \omega . \\
 \end{array}
\end{equation}
The quantities $\omega $ and $u_{1}$, which appear here on the right-hand
sides, are expressed in terms of $P$ and $M$ algebraically by the set of linear
equations
\begin{equation}
\label{eq6}
\begin{array}{c}
 u_1 + \mu (\dot {\xi }_p - \eta _p \omega ) = P / m, \\
 \left[ {(1 - \mu )a^2 + I_0 + \mu (\xi _p^2 + \eta _p^2 )} \right]\omega -
\mu \eta _p u_1 \\
= M / m - \mu (\xi _p \dot {\eta }_p - \dot {\xi }_p \eta _p). \\
 \end{array}
\end{equation}

Let $\xi _p = s$, $\eta _p = \varepsilon \sin \Omega t$. Using the
dimensionless time $\tau = \Omega t$ and the normalized variables
$p = P /m\Omega$, $q = M / m\Omega$, $u = u_1 / \Omega$, $w = \omega / \Omega$, we obtain
\begin{equation}
\label{eq7}
\begin{array}{l}
 \dot {p} = (Dw + \mu \varepsilon \cos \tau )w - \nu _1 u, \\
 \dot {q} = - (Dw + \mu \varepsilon \cos \tau )u - \nu _3 w, \\
 \end{array}
\end{equation}
where $D = a - \mu a + \mu s$, $\nu _1 = c_1 m^{ - 1}$, $\nu _3 = c_3 m^{- 1}$.
Now the dots denote derivatives with respect to the dimensionless
time $\tau $, and the formulae of constraints of the generalized velocities
and momenta take the form
\begin{equation}
\label{eq8}
\begin{array}{l}
 u - w\mu \varepsilon \sin \tau = p,\, \\
 - u\mu \varepsilon \sin \tau + (J + \mu \varepsilon ^2\sin ^2\tau )w = q -
\mu s\varepsilon \cos \tau , \\
 \end{array}
\end{equation}
where $J = I_0 + a^2 + \mu (s^2 - a^2)$. It is easy to verify that the
equations are invariant under the change of variables
\begin{equation}
\label{eq9}
\tau \to \tau + \pi ,\,\,\,q \to - q,\,p \to p,\,\,w \to - w,\,\,u \to u.
\end{equation}

Equations (\ref{eq7}) and (\ref{eq8}), which govern the evolution of two variables $p$, $q$ in
continuous time, define the reduced system, which can be considered
independently of the other variables relating to configuration space. These variables, namely, the coordinates of the point
where the knife edge $R$ is located in the laboratory reference frame $x$, $y$ and
the angle of rotation of the platform $\phi $, are governed by the equations
\begin{equation}
\label{eq10}
\dot {x} = u\cos \varphi ,\,\,\dot {y} = u\sin \varphi ,\,\,\dot {\varphi }
= \omega .
\end{equation}

This completes the mathematical formulation of the problem.

When analyzing the sleigh movements, information on the reaction force
acting at the point of application of the nonholonomic constraint
perpendicular to the ``knife edge'' may be useful. It is given just by the
Lagrange multiplier in the second equation (\ref{eq3}) and, as can be shown, in the
normalized form, is determined by the expression
\begin{equation}
\label{eq10a}
F = \lambda / m\Omega ^2 = uw - \varepsilon \mu (1 + w^2)\sin \tau + (a -
\mu a + \mu s)\dot {w}.
\end{equation}

Suppose that the values $p_n$, $q_n $ of the variables are assigned at
time $\tau =2\pi n$. Solving numerically the equations on the time interval
$\Delta \tau = 2\pi $ under these initial conditions, we can obtain the new values
\begin{equation}
\label{eq11}
(p_{n + 1} ,\,\,q_{n + 1} ) = {\rm {\bf f}}(p_n ,\,\,q_n ).
\end{equation}
Thus we have defined a two-dimensional stroboscopic Poincar\'{e} map, which
is convenient to describe and represent the results of analysis of the
dynamical behavior of the reduced system in discrete time.

The change of variable $q = z - p\mu \varepsilon \sin \tau $ brings the
reduced equations to an equivalent form, which is sometimes more convenient,
\begin{equation}
\label{eq12}
\begin{array}{l}
 \dot {p} = (Dw + \mu \varepsilon \cos \tau )w - \nu _1 u, \\
 \dot {z} = - (Dp + \nu _3 )w - \nu _1 u\mu \varepsilon \sin \tau , \\
 w = \frac{z - \mu s\varepsilon \cos \tau }{J + \mu (1 - \mu )\varepsilon
^2\sin ^2\tau },\,\,u = p + w\mu \varepsilon \sin \tau . \\
 \end{array}
\end{equation}

We note that in the case $\nu _{1}=0$ the terms containing the coefficient
$\nu _{3}$ are eliminated by a variable change $p = {p}' - \nu _3 D^{-1}$; therefore, at small friction, the rotational component of friction is,
generally speaking, of little importance.

\section{Acceleration under small oscillations}

Consider the motion of the sleigh in a case of small amplitude of
oscillations of a small internal mass using an analytical method inspired by
\cite{32} and developed for this specific problem in \cite{25}. Under the assumption
that the longitudinal momentum $p$ does not vary much over a period of
oscillations of the internal mass, we write for it an equation averaged over
a period:
\begin{equation}
\label{eq13}
\dot {p} = D\overline {w^2} + \mu \varepsilon \overline {w\cos \tau } - \nu
_1 \overline u .
\end{equation}

To find the averaged terms appearing on the right-hand side of this
expression, we neglect the quantity $\varepsilon ^{2}$ and, according to
the third equation of (\ref{eq12}), we set
$w = (z - \mu s\varepsilon \cos \tau )J^{- 1}$,
$u = p + J^{ - 1}\mu \varepsilon (z - \mu s\varepsilon \cos \tau )\sin \tau $. Substituting into the second equation of (\ref{eq12}) then gives
\begin{equation}
\label{eq14}
\dot {z} = J^{ - 1}(Dp + \nu _3 )( - z + \mu s\varepsilon \cos \tau ) - \nu
_1 p\mu \varepsilon \sin \tau ,
\end{equation}
and in the approximation of constancy of $p$ the solution has the form

\begin{equation}
\label{eq15}
\begin{array}{l}
 z = \mu \varepsilon \frac{(Dp + \nu _3 )^2s + \nu _1 pJ^2}{J^2 + (Dp + \nu
_3 )^2}\cos \tau \\
 \,\,\,\,\,\,\,\,\, + \mu \varepsilon J\frac{(Dp + \nu _3 )(s - \nu _1
p)}{J^2 + (Dp + \nu _3 )^2}\sin \tau . \\
 \end{array}
\end{equation}
\bigskip
It yields
\begin{equation}
\label{eq16}
\begin{array}{l}
 w = (z - \mu s\varepsilon \cos \tau )J^{ - 1} \\
 \,\,\,\,\,\,\,\,\,\,\,\,\,\,\,\,\, = \mu \varepsilon \frac{(\nu _1 p - s)(J\cos \tau - (Dp + \nu
_3 )\sin \tau )}{J^2 + (Dp + \nu _3 )^2},
 \end{array}
\end{equation}
\bigskip
so that
\begin{equation}
\label{eq17}
\begin{array}{l}
 \overline {w^2} = \textstyle{1 \over 2}\mu ^2\varepsilon ^2\frac{(s - \nu
_1 p)^2}{J^2 + (Dp + \nu _3 )^2},\\
  \overline u = p + \mu \varepsilon \overline {w\sin \tau } = p +
\textstyle{1 \over 2}\mu ^2\varepsilon ^2\frac{(Dp + \nu _3 )(s - \nu _1
p)}{J^2 + (Dp + \nu _3 )^2}, \\
 \overline {w\cos \tau } = \textstyle{1 \over 2}\mu \varepsilon
\frac{J(\nu _1 p - s)}{J^2 + (Dp + \nu _3 )^2}. \\
\end{array}
\end{equation}
\bigskip
Thus, equation (\ref{eq13}) takes the form
\begin{equation}
\label{eq18}
\begin{array}{l}
 \dot {p} + \nu _1 p = \\
 \textstyle{1 \over 2}\mu ^2\varepsilon ^2(s - \nu _1 p)\frac{Ds - J - (2Dp
+ \nu _3 )\nu _1 }{J^2 + (Dp + \nu _3 )^2}. \\
 \end{array}
\end{equation}
\bigskip
According to \cite{25}, under the conditions
\begin{equation}
\label{eq19}
\begin{array}{l}
 D = a - \mu a + \mu s > 0, \\
 s\left( {Ds - J} \right) = s\left( {a(s - a)(1 - \mu ) - I_0 } \right) > 0,
\\
 \end{array}
\end{equation}
in the absence of friction ($\nu _1 = \nu _3 = 0$) unbounded acceleration
occurs with an increase in the momentum of the sleigh. Indeed, integrating
equation (\ref{eq18}) in this case gives $\textstyle{1 \over 3}p^3D^2 + J^2p =
\textstyle{1 \over 2}\mu ^2s\varepsilon ^2(Ds - J)\tau + \mbox{const}$, and,
since in the region of a large momentum the first term on the left-hand side
dominates, the increase in the momentum (and of the velocity of the sleigh
in the direction of the knife edge) asymptotically follows the power law
$t^{1 / 3}$. In \cite{25}, a good agreement is demonstrated between analytical
results and those of numerical simulation of acceleration in the absence of
friction.

When friction is taken into account, the increase in the velocity ceases to
be unbounded and is stabilized at some level. This is illustrated in Fig.~2,
where for some specific parameters satisfying conditions (\ref{eq19}) the
longitudinal momentum is shown versus time according to the results of
numerical solution of equations (\ref{eq12}) (lines) and the averaged equation (\ref{eq18})
with zero initial conditions (dots). It can be seen that the agreement is
very good.

\begin{figure}[htbp]
\centerline{\includegraphics[width=3.3in]{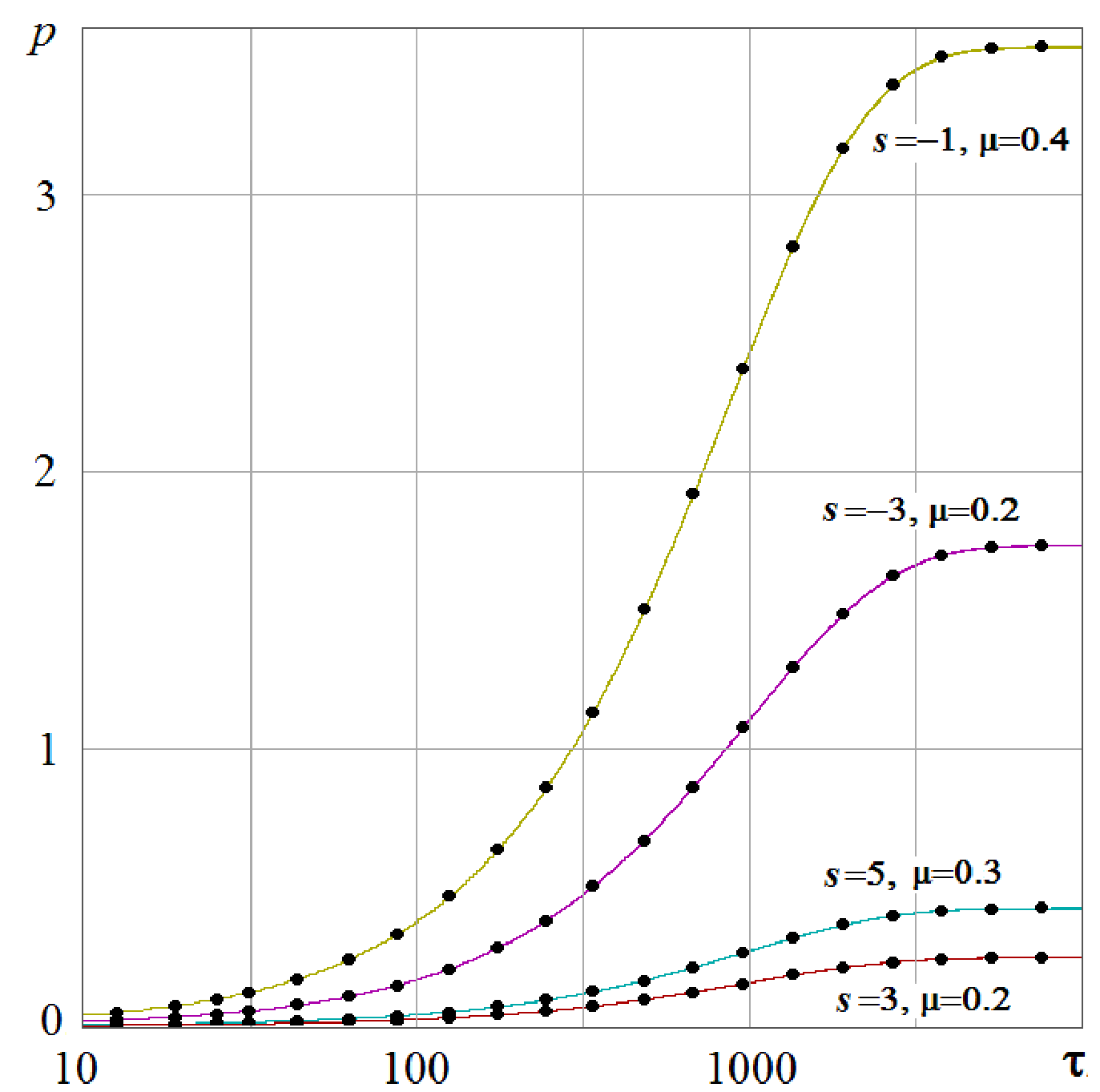}}
\label{fig2}
\caption{Dimensionless longitudinal momentum $p$ versus dimensionless time in
regions of accelerated motion of the sleigh for different values of $s$ and
$\mu $ as obtained from numerical solution of the equations (\ref{eq12}) (solid
lines) and from the shortened equation (\ref{eq18}) (dots). The other parameters are
$I_{0}=1$, $a=1$, $\varepsilon =0.3$, $\nu _{1}=0.001$, $\nu _{3}=0$
}
\end{figure}

Figure 3 shows the plane of the parameters $s$ (the coordinate of the
oscillating mass relative to the knife edge) and $\mu $ (the ratio of the
oscillating mass to the total mass of the system). White denotes regions
where conditions (\ref{eq19}) are satisfied, and inside these regions one can see
level lines (indicated by corresponding numbers) for the momentum values the
sleigh reaches within a prolonged time interval in the presence of friction.

\begin{figure}[htbp]
\centerline{\includegraphics[width=3.3in]{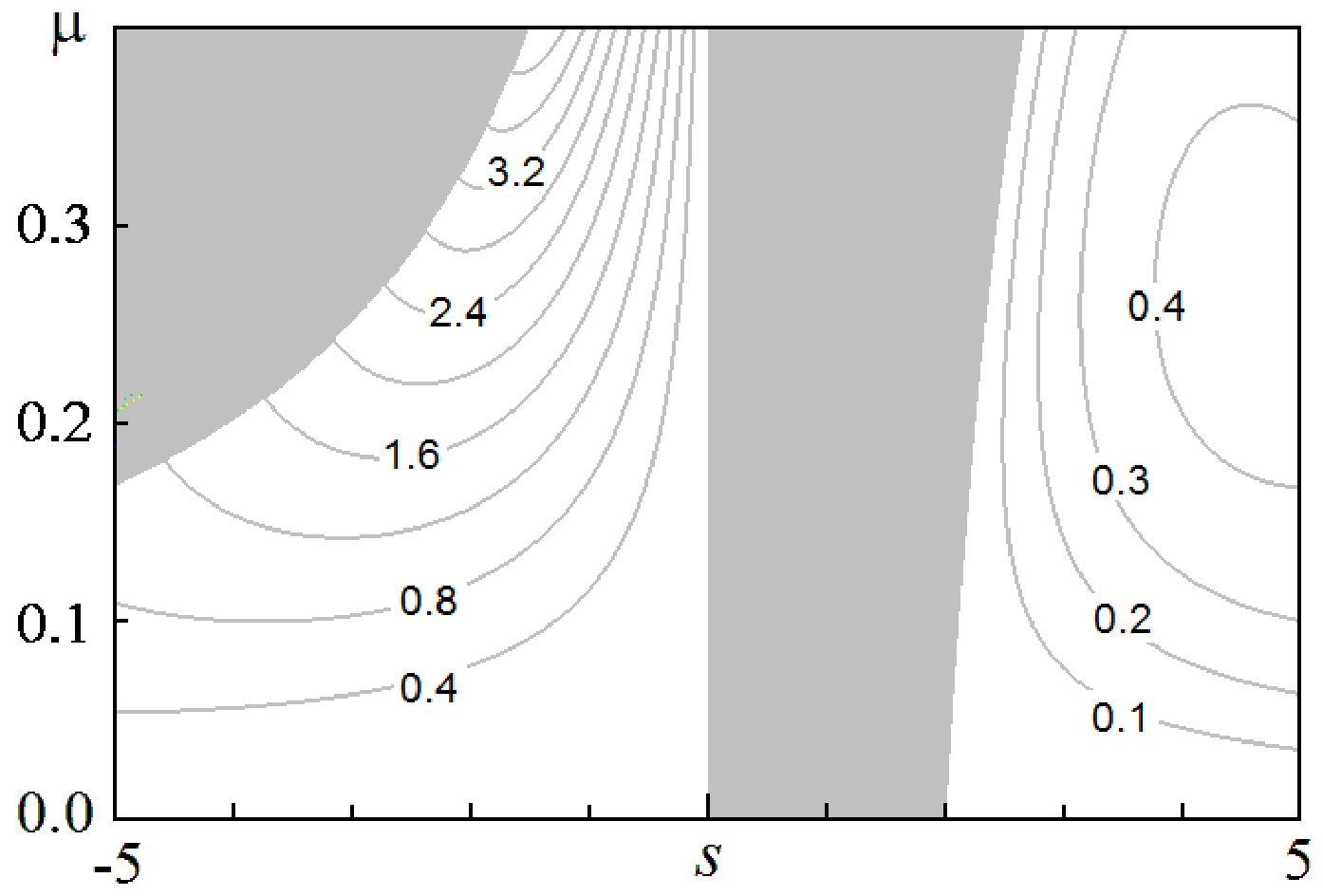}}
\label{fig3}
\caption{Regions of acceleration of the sleigh (white) and regions of
oscillatory variations in the velocity (grey) on the plane of parameters $s$
and $\mu $. The other parameters are $I_{0}=1$, $a=1$, $\varepsilon=0.3$,
$\nu_{1}=0.001$, $\nu _{3}=0$. In the region of acceleration, one can see
level lines of the function of the longitudinal momentum values, which the
sleigh reaches starting from a rest within a fairly large time interval
($10^{3}$ periods of oscillations of the internal mass).}
\end{figure}

We now will estimate the asymptotic characteristics of the motion of the
sleigh. For simplicity we assume that $\nu _{3}=0$, so that the equation
takes the form
\begin{equation}
\label{eq20}
\dot {p} = - \nu _1 p + \textstyle{1 \over 2}\mu ^2\varepsilon ^2(s - \nu _1
p)\frac{Ds - J - 2Dp\nu _1 }{J^2 + D^2p^2}.
\end{equation}

If $\mu ^2\varepsilon ^2 < < 1$, then, as the sleigh approaches steady
motion, which exists under conditions (\ref{eq19}) and corresponds to
$\nu _1 p\sim \mu ^2\varepsilon ^2$, one can neglect terms of higher order in the small
parameter and write
\begin{equation}
\label{eq21}
\dot {p} = - \nu _1 p + \textstyle{1 \over 2}\mu ^2\varepsilon ^2s\frac{Ds -
J}{J^2 + D^2p^2}.
\end{equation}

Thus, for the steady motion we have
\begin{equation}
\label{eq22}
pJ^2 + D^2p^3 = \textstyle{1 \over 2}\mu ^2\varepsilon ^2s(Ds - J)\nu _1^{ -
1} .
\end{equation}

As long as the parameter $\nu _{1}$ is not very small, the first term on
the left-hand side of (\ref{eq22}) dominates, and the momentum reached during
acceleration grows with decreasing coefficient of friction as a minus first
power:
$p\sim \textstyle{1 \over 2}\mu ^2\varepsilon ^2sJ^{ - 2}(Ds - J)\nu_1^{ - 1} $.
On the other hand, when $\nu _{1}$ is very small, the second
term on the left-hand side in (\ref{eq22}) dominates, and the momentum reached is
inversely proportional to the cubic root of the friction coefficient:
$p\sim \left( {\textstyle{1 \over 2}\mu ^2\varepsilon ^2sD^{ - 2}(Ds - J)}
\right)^{\raise0.5ex\hbox{$\scriptstyle
1$}\kern-0.1em/\kern-0.15em\lower0.25ex\hbox{$\scriptstyle 3$}}\nu _1^{ -
\raise0.5ex\hbox{$\scriptstyle
1$}\kern-0.1em/\kern-0.15em\lower0.25ex\hbox{$\scriptstyle 3$}} $. The
crossover from one type of dependence to another takes place near the value
of the parameter $\nu _1 \sim \mu ^2\varepsilon ^2s(Ds - J)DJ^{ - 3}$.

\section{Parametric mechanism of acceleration}

Consider a particular case of equations (\ref{eq12}) where the oscillating particle
moves in a straight line passing through the center of mass of the whole
system (platform plus particle). In this case, $D=0$, and the problem reduces
to a system of linear equations with variable coefficients
\begin{equation}
\label{eq23}
\begin{array}{l}
 \dot {p} = \mu \varepsilon w\cos \tau - \nu _1 u, \\
 \dot {z} = - \nu _3 w - \nu _1 u\mu \varepsilon \sin \tau , \\
 w = \frac{z - \mu s\varepsilon \cos \tau }{J + \mu (1 - \mu )\varepsilon
^2\sin ^2\tau },\\
u = p + w\mu \varepsilon \sin \tau ,
 \end{array}
\end{equation}
which is similar in structure to equations used in the linear theory of
parametric oscillations, such as the Mathieu equation. We note that the
presence of dissipation appears to be a necessary condition for the effect
of increasing parametric oscillations, since when there is absolutely no
friction ($\nu _{1}=\nu _{3}=0$), the second equation degenerates to
the trivial relation $\dot {z} = 0$, whence it follows that $z = const$.

For each period of the coefficient oscillations, the transformation of the
state is defined by a monodromy matrix, which can be found by numerically
integrating the homogeneous equations that correspond to the system (\ref{eq23})
\begin{equation}
\label{eq24}
\begin{array}{l}
 \dot {p} = \mu \varepsilon w\cos \tau - \nu _1 u,\,\, \\
 \dot {z} = - \nu _3 w - \nu _1 u\mu \varepsilon \sin \tau , \\
 w = z / [J + \mu (1 - \mu )\varepsilon ^2\sin ^2\tau ],\\
 u = p + w\mu \varepsilon \sin \tau . \\
 \end{array}
\end{equation}
in one period of oscillations with initial conditions (1, 0) and (0, 1),
which gives, respectively, the first and the second column of the matrix
\begin{equation}
\label{eq25}
\left( {{\begin{array}{*{20}c}
 {p_{n + 1} } \hfill \\
 {z_{n + 1} } \hfill \\
\end{array} }} \right) = \left( {{\begin{array}{*{20}c}
 {a_{11} } \hfill & {a_{12} } \hfill \\
 {a_{21} } \hfill & {a_{22} } \hfill \\
\end{array} }} \right)\left( {{\begin{array}{*{20}c}
 {p_n } \hfill \\
 {z_n } \hfill \\
\end{array} }} \right).
\end{equation}
\begin{figure}[htbp]
\centerline{\includegraphics[width=3.3in]{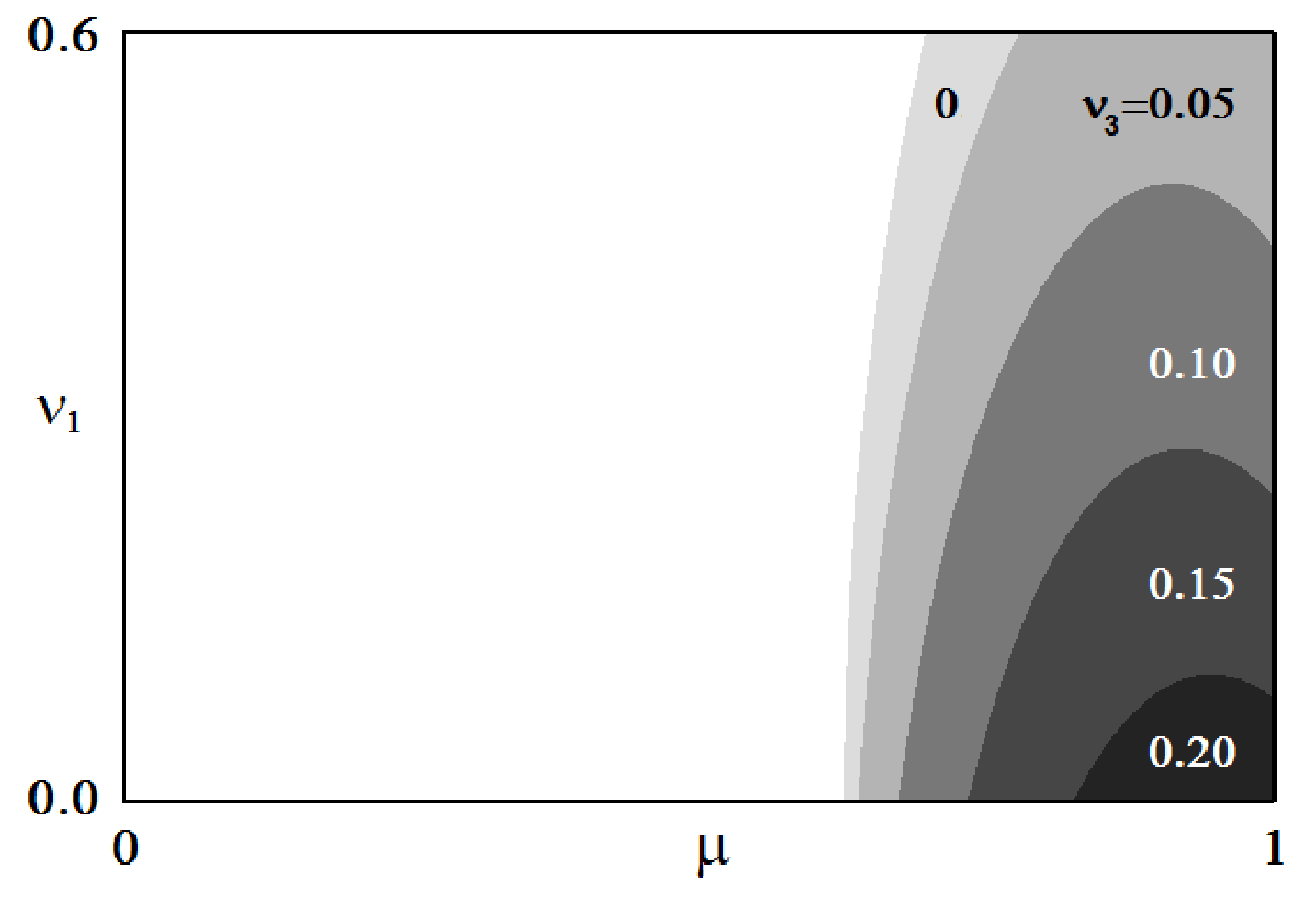}}
\label{fig4}
\caption{Regions of parametric instability on the plane of parameters (a
relative magnitude of the oscillating mass $\mu $ versus a coefficient of
friction for the translational motion $\nu _{1})$ for $J=0.15$, $\varepsilon=1$. The values of the friction coefficient for the rotational component of
motion $\nu _{3}$, which correspond to each region, are indicated by
numbers.}
\end{figure}

The condition for instability is that there be at least one eigenvalue of
the monodromy matrix whose absolute value exceeds 1. Figure 4 shows zones of
parametric instability for equations (\ref{eq12}) on the plane of parameters $\mu $
and $\nu _{1}$ for several values of $\nu _{3}$ (the coefficient of
friction of the rotational component of motion), which are indicated by
numbers. We note that, in contrast to the mechanism considered in the
previous section, acceleration in this case is achieved only when the
relative magnitude of the oscillating mass is large enough. Also, when the
parameters fit the zone of instability, acceleration to arbitrarily large
values of momentum takes place despite the energy losses due to dissipation
(friction). This is obvious from the linear nature of equations (\ref{eq23}) (see
also Fig.~5a).
It should be noted that such parametric acceleration is accompanied by an
unlimited increase in the oscillation of the reaction force, which
determines the motion of the sleigh and is given in the normalized form by
the expression (\ref{eq10a}). (The last term in this case vanishes.) In Fig.6 the
panel (a) shows the dependence of the reaction force versus time as obtained
in the numerical integration of the equations.

However, if the line of oscillations of the moving mass is displaced from
the center of mass, it leads to saturation of the parametric instability,
where the trajectory of the reduced equations evolves to an attractor. This
is illustrated by diagrams (b) and (c) of Fig.~5, in which a chaotic
attractor is approached. The degree of acceleration can be characterized by
the root-mean-square value of the longitudinal momentum on the attractor.
This value is bounded, but increases as the parameters approach the
degenerate situation of linearity of equations (\ref{eq23}). Now the oscillations of the reaction force appear also to be bounded, which
is illustrated by the panel (b) in Fig.6.

Chaotic dynamics on the attractor of the reduced system corresponds to
motion of the sleigh in the laboratory reference frame in the form of a
random two-dimensional walk of diffusion type, as is illustrated in Fig.~7.

Diagram (a) visualizes the results of numerical integration of equations for
coordinates (\ref{eq10}) together with the reduced equations (\ref{eq12}) and shows how the
trajectory of the sleigh looks like when the given parameters correspond to
Fig.5c. The sleigh is started from the origin of coordinates $x=0$, $y=0$ with
zero momentum $p$ and moment $q$ and with a zero initial angle of rotation
$\varphi $. The motion is tracked up to the instant where the distance from
the origin exceeds the given value $r_{\max } = 100$. Further calculations
continue with current values of $p$, $q$ and $\varphi $, but with a start again
from the origin. Thus, the figure shows several successive fragments of the
same trajectory, to which the orbit (approaching the chaotic attractor) of
the reduced equations in Fig.5c corresponds.

\begin{figure*}[htbp]
\centerline{\includegraphics[width=6.8in]{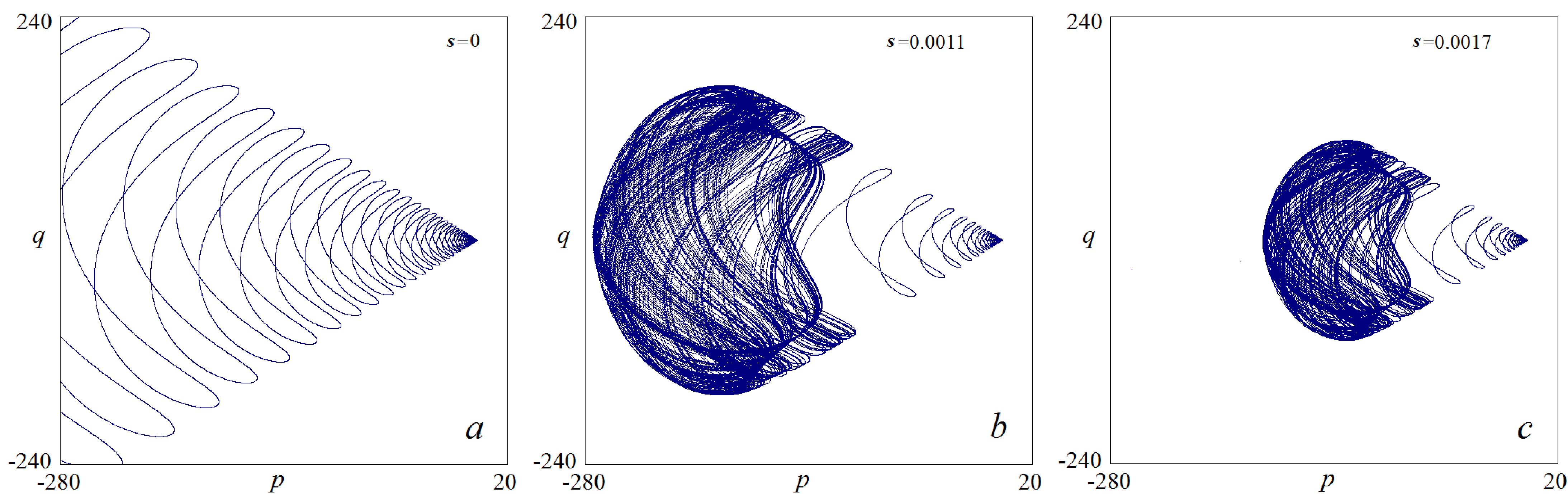}}
\label{fig5}
\caption{Unbounded acceleration of the sleigh in a situation where the
oscillating mass is located on the line passing through the center of mass
(a) and bounded acceleration where the trajectory approaches a chaotic
attractor (b), (c) with displacements from this line, according to the values of
$s$ indicated in the figure. The center of mass of the platform coincides with
the position of the knife edge, $a=0$. The other parameters are:
$I_0 =0.15$, $\mu = 0.85$, $\nu _1 = 0.35$, $\nu _3 = 0.1$, $\varepsilon = 1$.}
\end{figure*}

\begin{figure*}[htbp]
\centerline{\includegraphics[width=6.8in]{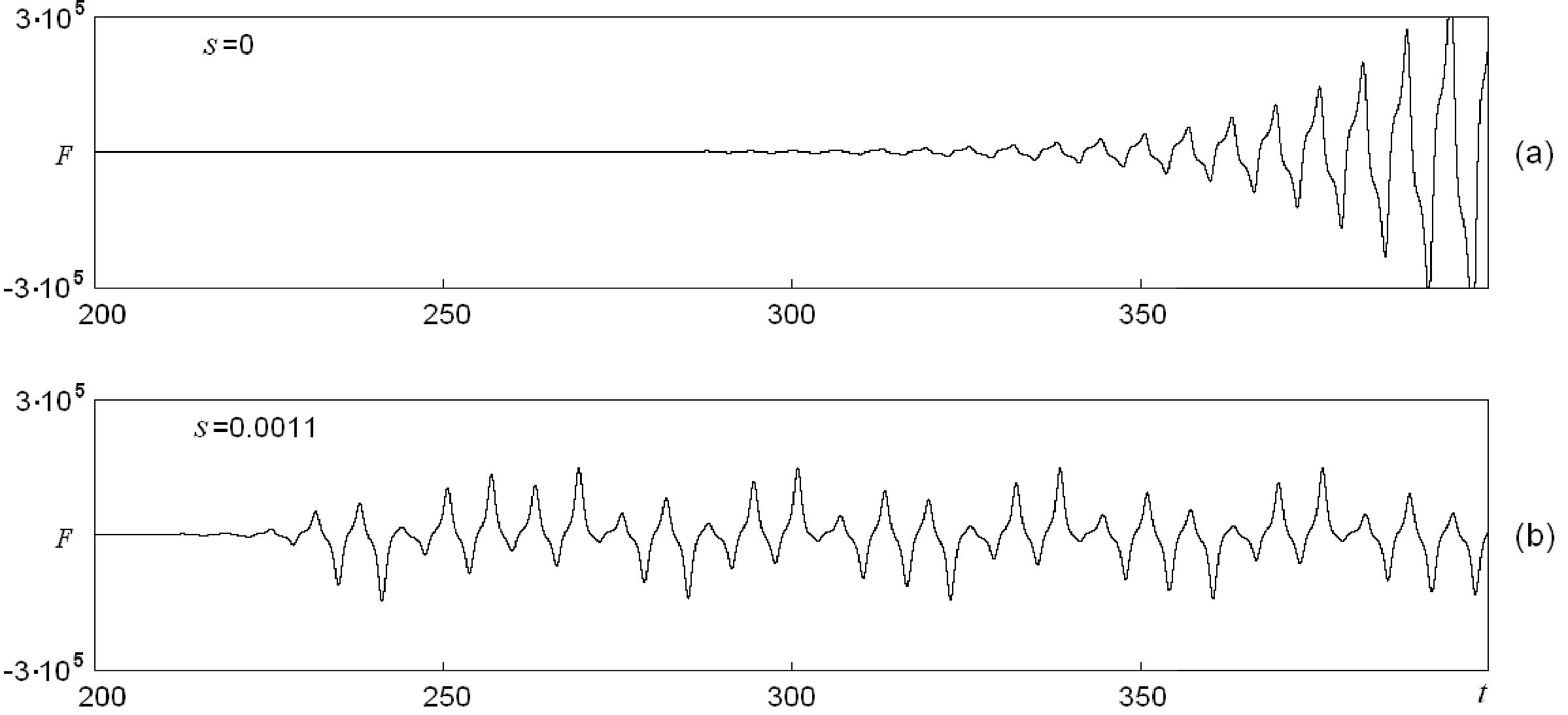}}
\label{fig6}
\caption{Dependence of the normalized reaction force $F = \lambda / m\Omega
^2$ acting at the location of the non-holonomic constraint versus time in
the situation of unlimited acceleration at $s$=0 (a), and in the case of
bounded acceleration with approach to the chaotic attractor in the phase
space of the reduced equations at $s$=0.0011 (b). The center of mass of the
platform coincides with the position of the knife-edge, $a$=0. Other parameters
are $I_0 = 0.15,\,\,\mu = 0.85,\,\nu _1 = 0.35,\,\nu _3 =
0.1,\,\,\varepsilon = 1$.}
\end{figure*}

It can be seen that the motion observed in the case at hand is a
two-dimensional random walk without a preferred direction, with the loss of
memory of the initial sleigh orientation for large time scales \cite{33,34,35}.
In this situation, the distribution of distances from the starting point to
the end point of the trajectory segment travelled over a fixed number of
oscillations of the internal mass $N$ (the number of iterations of the
Poincar\'{e} map) must asymptotically tend to the Rayleigh distribution, and
the distribution of azimuth angles, to a uniform distribution in the
interval $[0,\,2\pi )$.

Diagrams (b) and (c) show cumulative distributions for the distances
$r =\sqrt {(x_{(k + 1)N} - x_{kN} )^2 + (y_{(k + 1)N} - y_{kN} )^2} $ and the
angles
$\theta = \arg [(x_{(k + 1)N} - x_{kN} ) + i(y_{(k + 1)N} - y_{kN})]$.
The solid lines correspond to numerical results for a specified number
of periods, $N=1000$. The set whose elements are fragments of the
above-mentioned reference trajectory, which are labeled with index
$k=1 \ldots 10^{4}$, plays the role of the sample space. The dots in panel (a)
correspond to the Rayleigh distribution
$F(r) = 1 - e^{ - r^2 / 2\sigma ^2}$, where $2\sigma ^2$ is the sum of dispersions of random values $x_{(k + 1)N} - x_{kN} $ and $y_{(k + 1)N} - y_{kN} $, which have been obtained as
statistical estimates from sampling; in the case at hand we have $\sigma ^2
\approx 5.3 \cdot 10^5$.

\begin{figure*}[htbp]
\centerline{\includegraphics[width=6.8in]{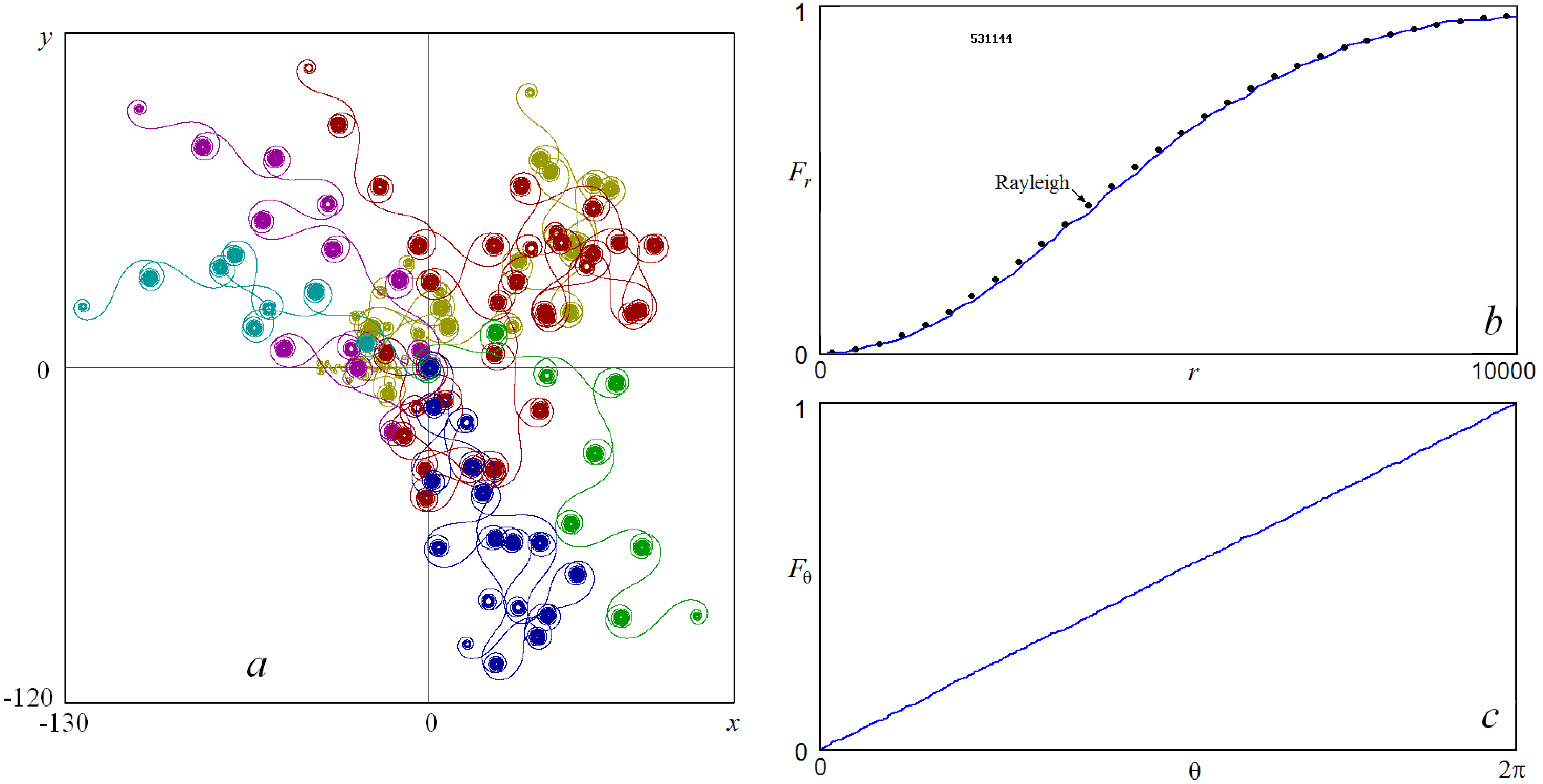}}
\label{fig7}
\caption{Family of fragments of the trajectory of the sleigh in the laboratory
reference frame, visualized as described in the text (a) and cumulative
functions of distribution for the distances travelled (b) and for the
azimuth angles (c), which were obtained for $10^{3}$ iterations of the
Poincar\'{e} map. The data presented in diagram (b) are compared with the
Rayleigh distribution (dots), and diagram (c) is indicative of a uniform
distribution of the angles}
\end{figure*}

Generally speaking, attractor of the reduced system, which is approached
under the conditions of a bounded parametric acceleration, can be chaotic or
regular.
\begin{figure*}[htbp]
\centerline{\includegraphics[width=6.8in]{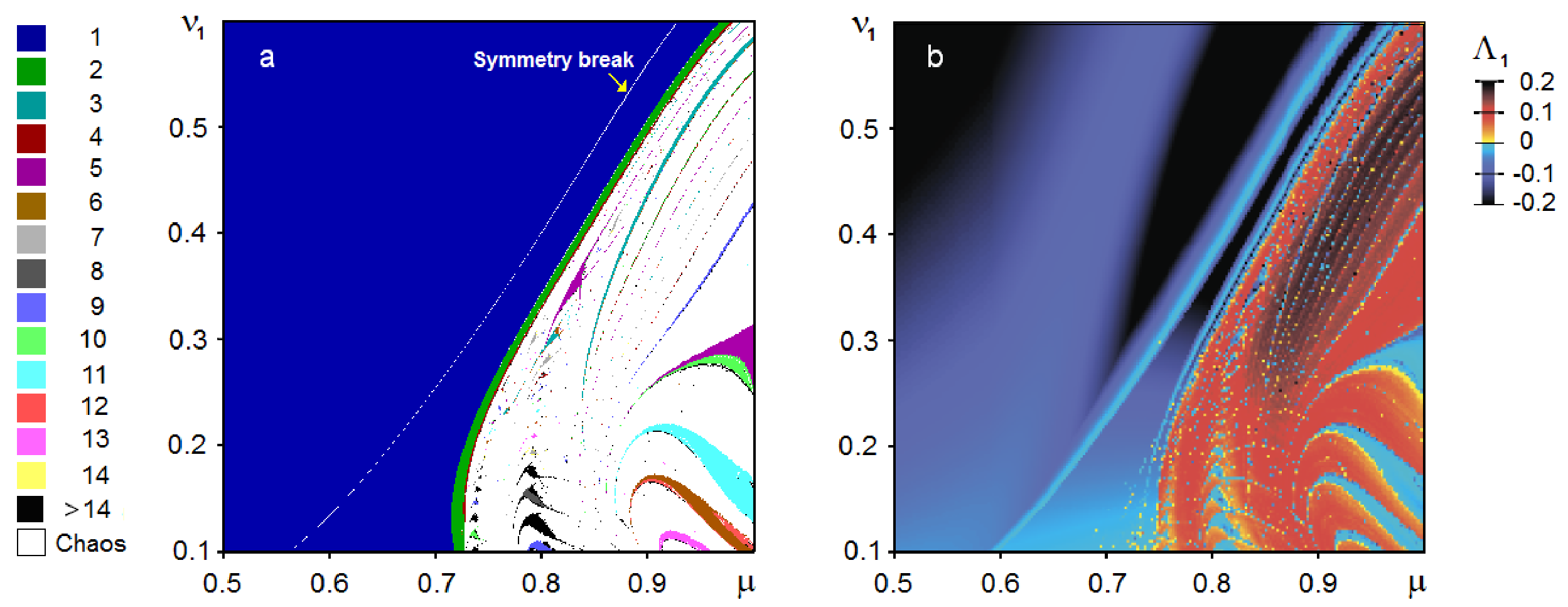}}
\label{fig7}
\caption{Chart of periodic and chaotic regimes on the plane of parameters
($\mu$, $\nu _{1}$) (a) and chart of Lyapunov exponents (b). The other
parameters are: $I_0 = 0.15$, $\nu _3 = 0.1$, $\varepsilon = 1$, $a = 0$, $s = 0.1$}
\end{figure*}
\begin{figure*}[htbp]
\centerline{\includegraphics[width=5.1in]{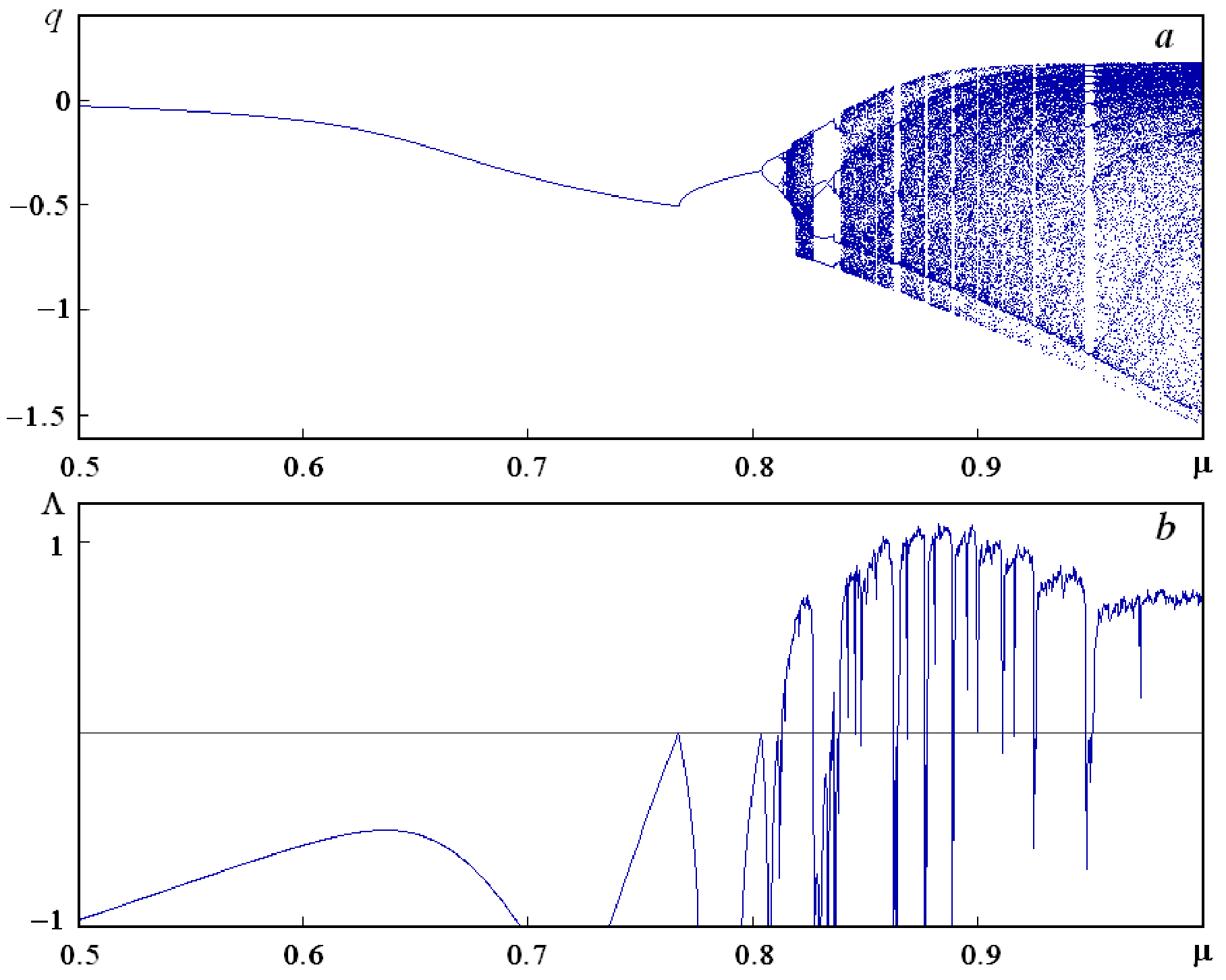}}
\label{fig9}
\caption{The bifurcation diagram showing the dynamical variable $q$ on attractor
of the Poincar\'{e} map versus parameter $\mu $ (a), and the plot of the
largest Lyapunov exponent (b). The other parameters are: $I_0 = 0.15$,
$\nu_3 = 0.1$, $\varepsilon = 1$, $a = 0$, $s = 0.1$}
\end{figure*}
Figure~8 shows two charts, plotted using different techniques, on the
parameter plane (the same as in Fig.~4, where the regions of parametric
instability were represented). The left panel presents a chart of regimes,
where periodic ones are indicated with colors, and white denotes the region
of chaos in accordance with the legend shown on the left. The right panel
shows a chart of the same part of the parameter plane (plotted by analogy
with a geographic map), where the colors code levels of the largest Lyapunov
exponent: the levels below zero are indicated in blue, and the levels above
zero are shown in yellow, red and brown, see the legend on the right of the
chart.

Figure~9 contains a bifurcation diagram showing values of the dynamical
variable $q$ (dimensionless angular momentum), which correspond to the attractor
of the Poincar\'{e} map, versus the parameter of relative magnitude of the
oscillating mass $\mu$, and a graph of the largest Lyapunov exponent in the
same interval of the parameter. On the parameter plane of Fig.~8, these
diagrams correspond to the horizontal midline for $\nu _{1}=0.35$. Figure~10 depicts the attractors of the system with continuous time at
representative points on the axis of $\mu $. These attractors are shown in
light blue, and the dots corresponding to the stroboscopic Poincar\'{e}
section are shown in red. Each panel contains an indication of the Lyapunov
exponents for the Poincar\'{e} map which have been found numerically for the
corresponding attractor.

When one moves along the axis of parameter $\mu $ from left to right, one
first observes an attractor in the form of a symmetric loop, which
corresponds to an attractive fixed point of the Poincar\'{e} map (panel
(a)). Then one observes a symmetry-loss bifurcation ($\mu \approx 0.765$),
followed by a cascade of period-doubling bifurcations (panels (b) and (c)),
and a transition to chaos. The resulting chaotic attractor is first
asymmetric (d), so that in the phase space it coexists with a symmetric
attractor resulting from the substitution (\ref{eq10}). As the parameter $\mu $
grows, the attractor and its partner merge to form a single symmetric
attractor (e). Such symmetric attractors can be seen in panels (g) and (i);
on the other hand, panel (h) demonstrates a chaotic asymmetric attractor.
The region of chaos is interspersed with periodicity windows, which look
like light-colored vertical strips in the bifurcation diagram and like dips
into a negative region in the graph of Lyapunov exponent. The attractor
shown in panel (f) just corresponds to one of the regularity windows.

\begin{figure*}[htbp]
\centerline{\includegraphics[width=6in]{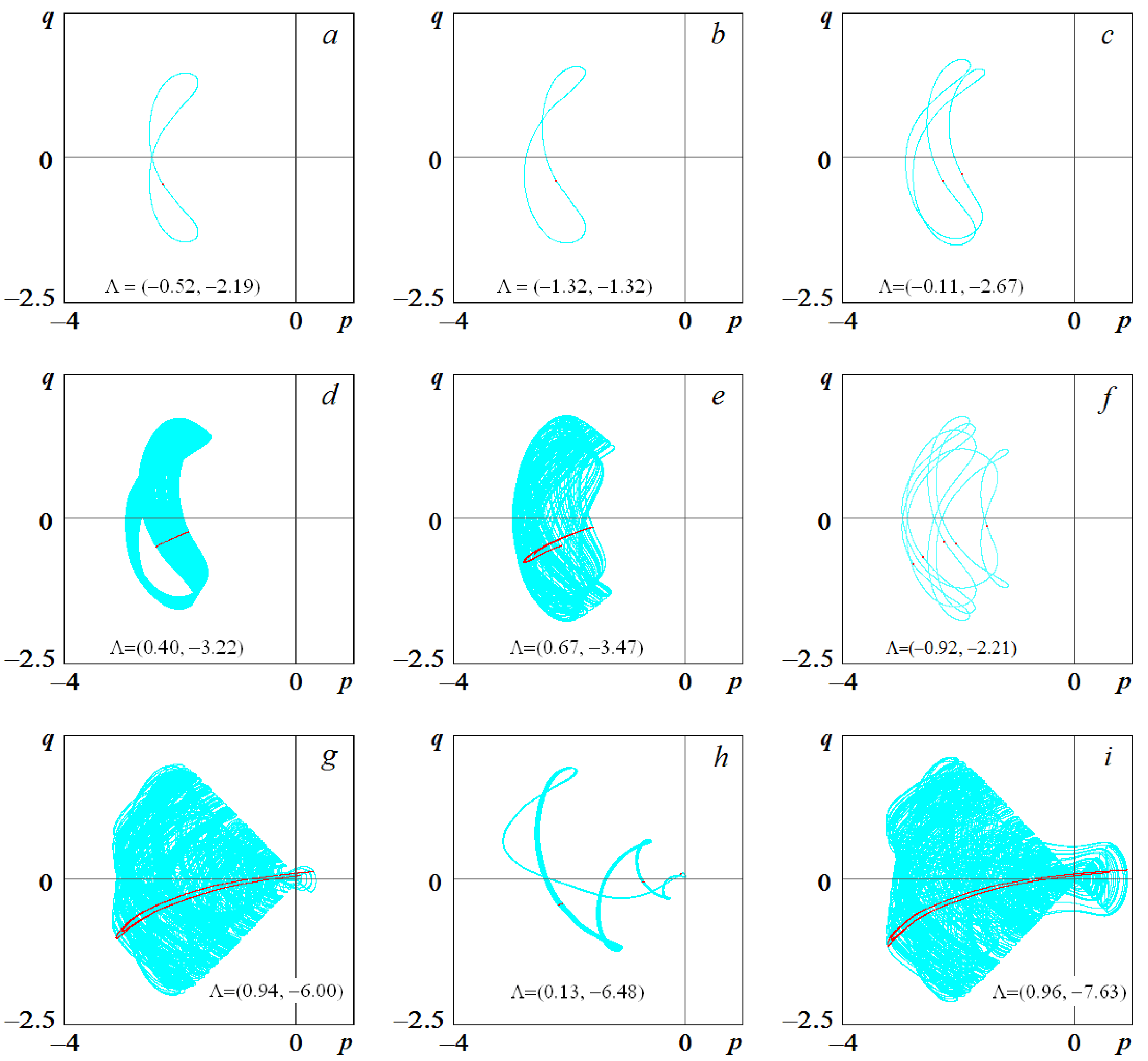}}
\label{fig10}
\caption{Attractors of the reduced equations for parameters
$I_0 =0.15$, $\nu _3 = 0.1$, $\varepsilon = 1$, $a = 0$, $s = 0.1$ and
$\mu=0.75$ (a), $0.78$ (b), $0.81$ (c), $0.816$ (d), $0.825$ (e), $0.83$ (f), $0.89$ (g),
$0.9$ (h), $0.92$ (i). Light-blue denotes trajectories in continuous time, and
red indicates dots corresponding to the stroboscopic Poincar\'{e} section at
time instants $t=2\pi n$. The Lyapunov exponents of the stroboscopic map are
indicated under the portraits of the attractors}
\end{figure*}

\section{Comparison of chaotic dynamics with and without friction}
The object we consider here, namely, the Chaplygin sleigh with an
oscillating internal mass, in the absence of friction is an example of
behavior specific to nonholonomic systems with complex dynamics, which
occurs due to invariance under time reversal. This implies a possibility of
coexistence in the phase space of objects typical for conservative systems
(regions occupied by closed invariant curves similar to KAM tori and regions
of chaos -- ``chaotic seas'') and typical for dissipative systems (regular
and chaotic attractors). Due to invariance under time reversal, each
attractor corresponds to a symmetric partner coexisting with it in the phase
space -- a repeller to which the phase trajectories are attracted when the
dynamics is tracked in backward time. When friction is incorporated, these
features, specific to reversible dynamics, degrade, and the behavior of the
system acquires features typical of usual dissipative dynamics. In the
concluding section of this paper we discuss the corresponding change in the
dynamical behavior of our system, and the effect it has on the properties of
the observed attractors.
\begin{figure*}[htbp]
\centerline{\includegraphics[width=5in]{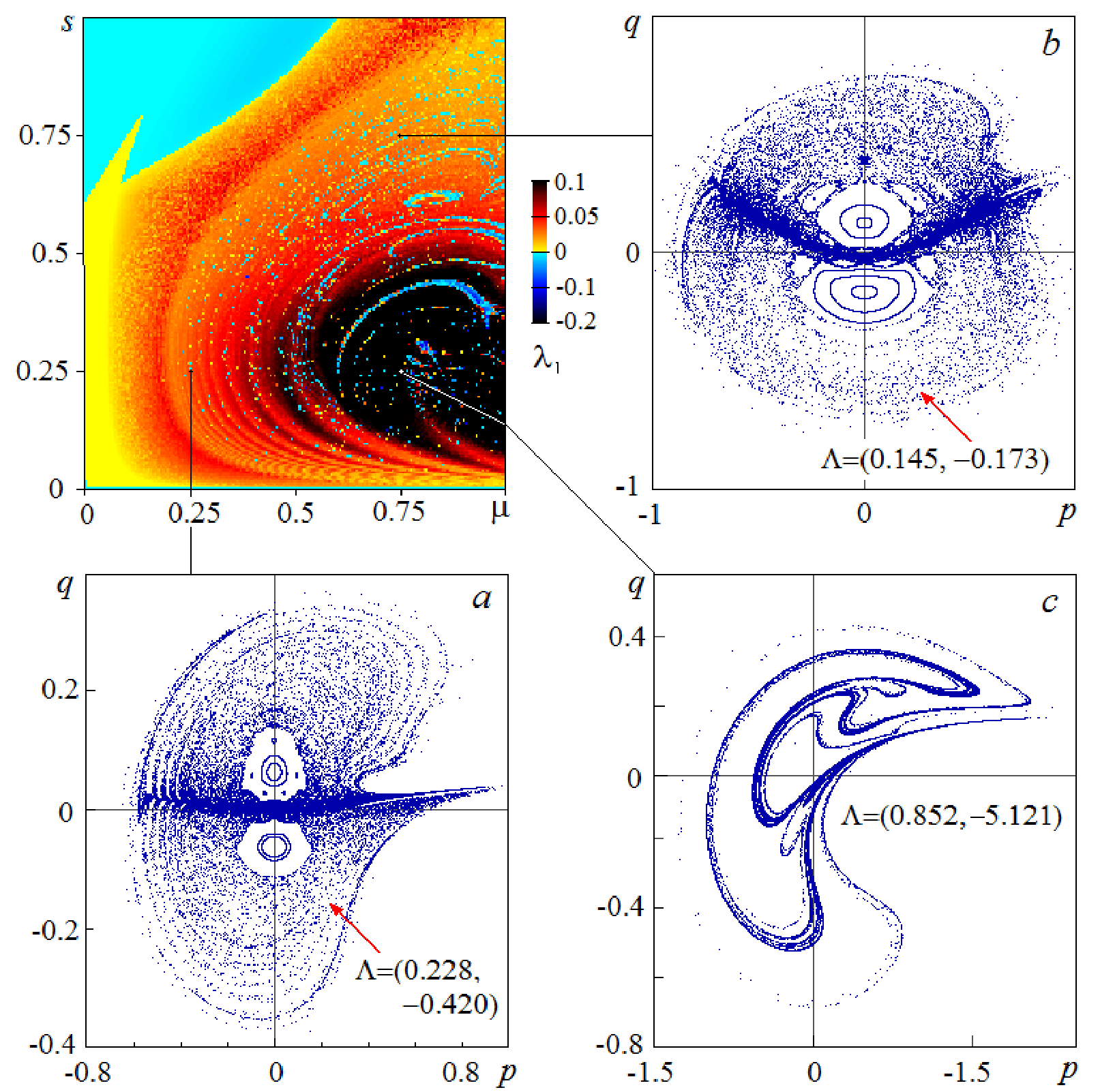}}
\label{fig10}
\caption{Chart of Lyapunov exponents on the parameter plane for a system
without dissipation and portraits illustrating the structure of the phase
plane in the regions where it has a similarity with conservative dynamics
(a), (b) and with dissipative dynamics (c). The values of the parameters are
$\mu =0.25$, $s=0.25$ (a), $\mu =0.25$, $s=0.75$ (b), $\mu =0.75$, $s=0.25$ (c);
the values of the other parameters are $\varepsilon =1$, $a=0.5$,
$I_{0}=0.05$, $\nu _{1}=0$, $\nu _{3}=0$. The Lyapunov exponents of the
chaotic dynamics for the Poincar\'{e} map are indicated in panels (a)-(c)}
\end{figure*}

In the absence of friction, when $\nu _{1}= 0$, $\nu _{3}= 0$, equations
(\ref{eq7})-(\ref{eq8}) take the form
\begin{equation}
\label{eq26}
\begin{array}{l}
 \dot {p} = (Dw + \mu \varepsilon \cos \tau )w,\,\,\,\,\, \\
 \dot {q} = - (Dw + \mu \varepsilon \cos \tau )u, \\
 u - w\mu \varepsilon \sin \tau = p,\\
 - u\mu \varepsilon \sin \tau + (J
+ \mu \varepsilon ^2\sin ^2\tau )w = q - \mu s\varepsilon \cos \tau , \\
 \end{array}
\end{equation}
and in this case, along with the symmetry (\ref{eq9}), invariance under time
reversal takes place. This corresponds to the involution
\begin{equation}
\label{eq27}
p \to - p,\,\,\,q \to q,\,\,\tau \to - \tau ,\,\,u \to - u,\,w \to w.
\end{equation}
Thus, if the system has an attractor, then a repeller symmetric relative to
the above involution will necessarily coexist with this attractor in the
phase space.

The chart in Fig.~11 gives an idea of the parameter space structure of the
reversible system: the Lyapunov exponent that determines the coloring of
each pixel according to the legend was calculated for the corresponding
parameters and identical initial conditions ($p=5$, $q=0$, $\tau =0$). At the
periphery one can see phase portraits of the stroboscopic Poincar\'{e} map
at representative points (a), (b) and (c), which were plotted in each case
for several sets of initial conditions.

The structure of the phase plane shown in diagrams (a) and (b) is similar to
that for conservative Hamiltonian systems with discrete time, namely, the
regions occupied by closed invariant curves similar to KAM tori, and the
chaotic sea regions. The latter should be treated as a sort of ``fat
attractors'', since the positive and negative Lyapunov exponents for them
differ in absolute value (see the values of the exponents for the
Poincar\'{e} map which are indicated in the panels). At the same time, for
motions on invariant curves the spectrum of Lyapunov exponents according to
calculations consists of two zero exponents (with accuracy to calculation
errors).

In diagram (c), one can see an attractor having a pronounced transverse
Cantor-like structure of filaments similar to attractors in dissipative
systems, while one observes no regions occupied by invariant curves.

We note that the estimate of the dimension from the Kaplan-Yorke formula
$D_{KY} = 1 + \Lambda _1 / \vert \Lambda _2 \vert $ for the ``fat''
attractors (a) and (b) gives 1.54 and 1.84, i.e., the dimension is rather
large. For the ``thin'' quasi-dissipative attractor (c) the dimension is
only 1.17.

\begin{figure*}[htbp]
\centerline{\includegraphics[width=6.8in]{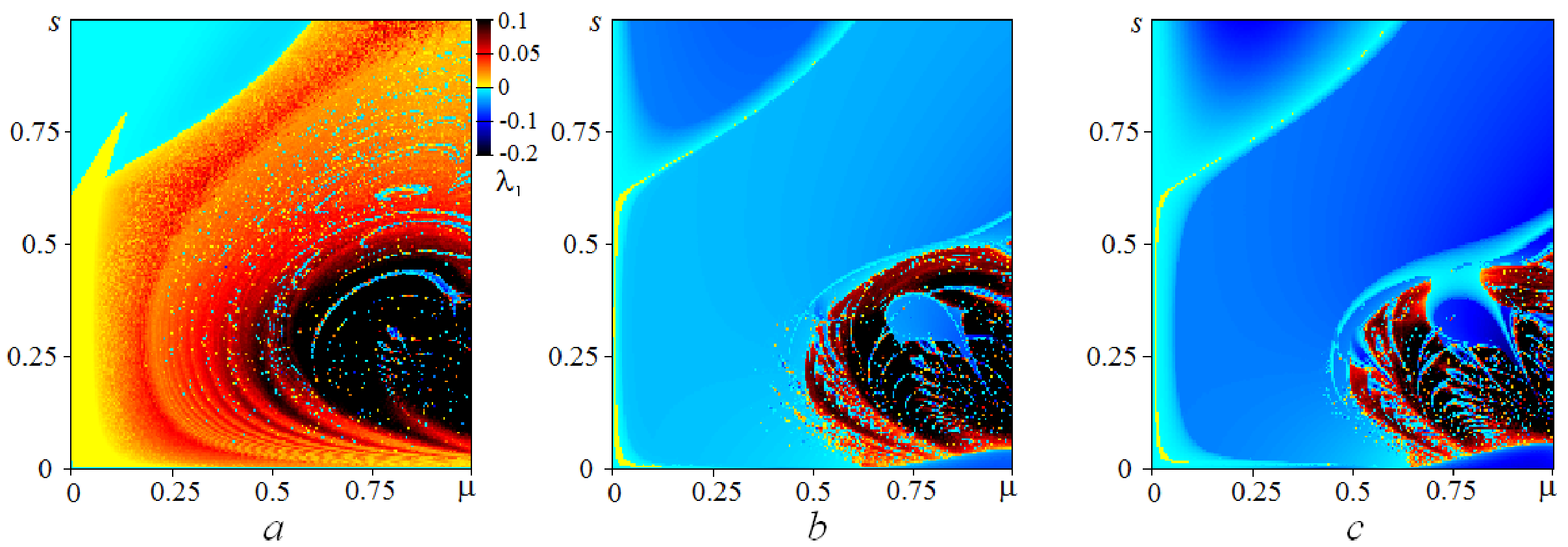}}
\label{fig12}
\caption{Lyapunov charts for a system without friction $\nu _{1}=0$ (a) and
with friction coefficients $\nu _{1}=0.05$ (b) and $0.1$ (c). The other
parameters are: $\varepsilon =1$, $a=0.5$, $I_{0}=0.05$, $\nu _{3}=0$}
\end{figure*}

\begin{figure*}[htbp]
\centerline{\includegraphics[width=6.6in]{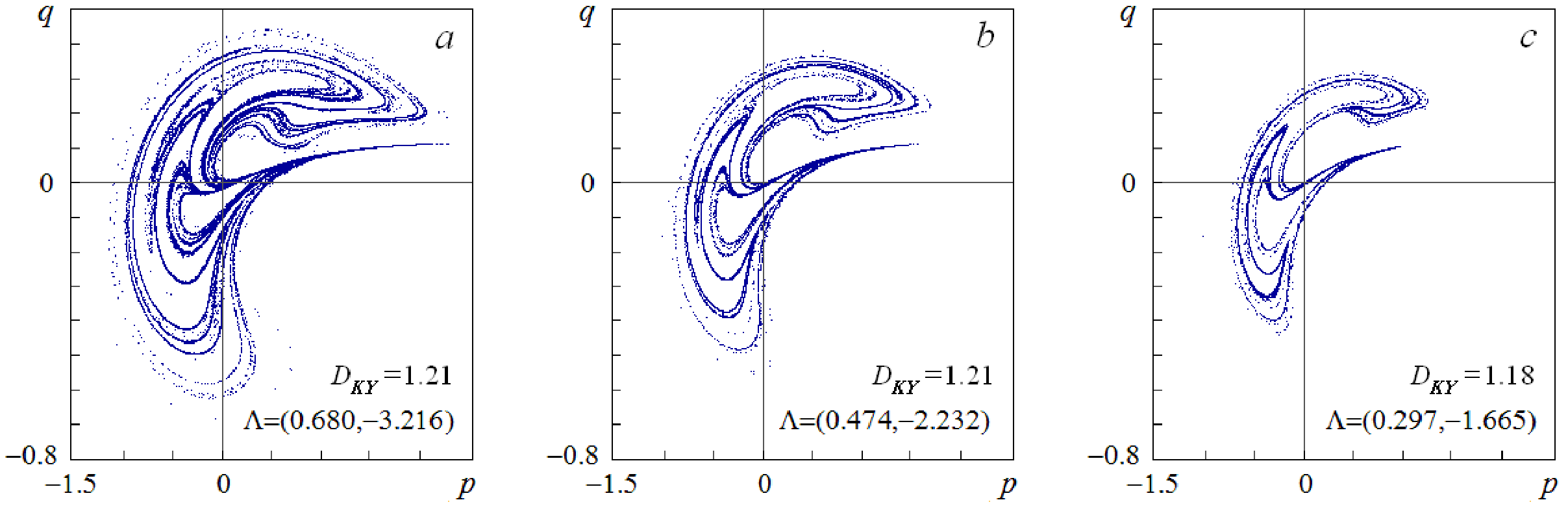}}
\label{fig13}
\caption{Portraits of attractors of the stroboscopic Poincar\'{e} map for a
system without friction $\nu _{1}=0$ (a) and for that with friction
coefficients $\nu _{1}=0.05$ (b) and $0.1$ (c). The other parameters are:
$\varepsilon =1$, $a=0.5$, $I_{0}=0.05$, $s=0.2$, $\mu =0.6$, $\nu _{3}=0$. In
the lower parts of the panels, Lyapunov exponents for the attractor of the
Poincar\'{e} map are indicated and the estimates of the Kaplan-Yorke
dimension are given}
\end{figure*}

Figure~12 presents a comparison of the Lyapunov charts for a system without
friction $\nu _{1}=0$ (a) and for friction coefficients $\nu _{1}=0.05$
(b) and 0.1 (c). In all cases it is assumed that $\nu _{3}=0$. As can be
seen, introducing even a small dissipation leads to ``washing-out'' of the
region of existence of chaos associated with fat attractors. These are now
occupied by regimes of regular dynamics. In contrast, there is a relatively
small influence of dissipation on chaotic attractors of the type shown in
Fig.~11c, which reside in the black region in the right lower part of the
chart. The action of friction reduces to some changes in the quantitative
characteristics (Lyapunov exponents), however, the general structure of the
attractors remains approximately the same, moreover, the Kaplan-Yorke
dimension does almost not change over the above-mentioned range of values of
the dissipation parameter. The aforesaid is illustrated by portraits of
attractors in the stroboscopic section in Fig.~13, which have been
constructed for $\nu _{1}=0$ (a), 0.05 (b) and 0.1 (c). In the lower parts
of the panels, Lyapunov exponents for the Poincar\'{e} map are indicated and
estimates of the Kaplan-York dimension are given.

Thus, it may be concluded that chaotic motions of a nonholonomic system are
divided into two classes and occur in specific parameter regions.

One kind of chaos is associated with fat attractors, which can be compared
with chaotic seas of conservative Hamiltonian systems. Like chaotic seas,
fat attractors coexist with phase space regions which are filled with
invariant curves and correspond to quasi-periodic dynamics. In the chart of
Fig.~11, such dynamics correspond to the orange-red region.

On the other hand, the chaos related to attractors of the second kind is
analogous to dissipative systems where the attractor exhibits a pronounced
transverse split structure of filaments. This corresponds to the dark region
in the right lower part of the chart in Fig.~11 (upper left panel).

The difference between both types of attractors lays in the presence or
absence of a structure in the form of split filaments and thus in dimension
properties -- for fat attractors the dimension is much higher. The
difference also lays in the degree of sensitivity to incorporation of
friction: in the region where fat attractors manifest themselves, chaos is
effectively washed out when dissipation is incorporated, and is replaced
with regular motions. This does not happen to attractors of the second kind,
and chaos survives in the presence of dissipation, at least as long as the
latter does not exceed some moderate level. Thus, in the presence of
friction, the specific properties of the nonholonomic system become less
pronounced, and the behavior becomes to a greater extent similar to that
typical of dissipative systems with complex dynamics.

\section{Conclusion}

The motion of the Chaplygin sleigh in the presence of an oscillating
internal mass in the presence of viscous friction is considered. It is
assumed that a material point oscillates relative to the main platform when
moving in a straight line perpendicularly to the axis passing through the
center of mass of the platform and through the point of application of the
nonholonomic constraint in the direction coinciding with that of the knife
edge which ensures the nonholonomic constraint. Equations governing the
dynamics of the system are formulated. The time evolution of momentum and
angular momentum is governed by a reduced system of equations (decoupled
from the other equations pertaining to the configuration space), and during
the motion of the internal mass their change is determined by a
two-dimensional map. The other configuration variables, including the
coordinates in the laboratory reference frame and the angle of rotation of
the sleigh, are expressed in terms of the momentum variables using
quadratures.

The results presented in this paper can be extended to wheeled vehicles,
since the nonholonomic constraint corresponding to the Chaplygin sleigh is
equivalent to that which is introduced by replacing the knife edge by a
wheel pair \cite{17} with the other supports sliding freely.

For the motion of the Chaplygin sleigh in the presence of weak friction, two
mechanisms of acceleration due to oscillations of the internal mass have
been found and investigated. Under small oscillations of the internal mass
the sleigh may exhibit acceleration of motion that is on average
rectilinear; in the absence of friction the acceleration is unbounded, while
in the presence of friction it becomes modified so that the velocity reached
is stabilized at a fixed level -- the smaller the friction, the higher the
velocity. The other mechanism is due to the effect of parametric excitation
of oscillations, when the mass which executes oscillations is comparable in
magnitude to the mass of the main platform. In this case, the presence of
friction is necessary for acceleration. In the simplest case of motion of
the internal mass in a straight line passing through the center of mass, the
problem reduces to a linear equation with periodic coefficients, and the
increase in parametric oscillations is unbounded, as in the Mathieu
equation. As is usually the case under parametric oscillations, linear
dissipation does not lead to saturation; the saturation occurs only in the
presence of nonlinear effects. In our case, the parametric instability and
the resulting acceleration of the sleigh are bounded if the line of
oscillations of the moving mass is displaced from the center of mass. The
sustained regime of motion is associated with the attractor of the reduced
system of equations, which in many cases turns out to be chaotic;
accordingly, the motion of the sleigh is also chaotic and similar to a
diffusion process.

Also, a qualitative analysis has been made of the dynamics in the absence of
friction, when the model under consideration provides an example of complex
dynamics which is specific to nonholonomic systems and is due to invariance
under time reversal, and a modification of this behavior with dissipation
has been discussed. In particular, it is shown that the incorporation of
dissipation drastically changes the dynamics in parameter regions where the
model without friction demonstrates coexistence of quasi-conservative types
of behavior, namely, phase space regions filled with invariant curves and
chaos corresponding to fat attractors. On the other hand, in the parameter
region where there are attractors similar in nature to attractors of
dissipative systems (fractal transverse structure of filaments), the
introduction of dissipation leads only to certain quantitative changes in
the spectrum of Lyapunov exponents, without a considerable change in the
dimension of the attractors.

The existence of chaotic attractors exhibited by the Chaplygin sleigh with a
moving internal mass makes it possible to apply chaos control methods \cite{36}
for organizing the motion in a specific direction by changing the
characteristics of oscillatory motion of the internal mass under conditions
of diffusion type sleigh dynamics. Since chaos is sensitive to small
perturbations, control can be performed by an arbitrarily small targeted
action.

The results obtained here supplement the methods (considered in other
publications) for controlling the motions of mobile mechanical systems whose
characteristics can be changed by regulating the parameters, including the
coefficient of intensity of impulses of external forces \cite{19,20}, by
switching the nonholonomic constraint to different locations \cite{22,23}, and
by changing the amplitude of oscillations of the internal mass or its
position \cite{24,25}.

\begin{acknowledgements}
This work was supported by grant No. 15-12-20035 of the Russian Science Foundation.
\end{acknowledgements}

\end{document}